\documentclass[12pt]{iopart}
\usepackage{graphicx}
\usepackage{subfigure}

\begin{document}

\title{Theory of magnetic phases of hexagonal rare earth manganites}

\author{I. Munawar and S. H. Curnoe}
\address{Department of Physics and Physical Oceanography,
Memorial University of Newfoundland, St.\ John's, Newfoundland \& Labrador 
A1B 3X7, Canada}
\ead{curnoe@physics.mun.ca}

\begin{abstract}

The magnetic phases of  hexagonal perovskites
RMnO$_3$ (R=Ho, Er, Tm, Yb, Sc, Y) are analysed using group theory and the
Landau
theory of phase transitions.
The competition between various magnetic order parameters
is discussed in the context of antiferromagnetic interactions.
A phenomenological model based on four one-dimensional magnetic order parameters is developed and studied numerically. 
It is shown that coupling of the various order parameters leads to a 
complex magnetic field-temperature phase diagram
and the results are compared to experiment.

\end{abstract}



\section{Introduction}
Hexagonal perovskites
RMnO$_3$ (R= Ho, Er, Tm, Yb, Lu, Y or Sc) belong to
an unusual class of materials known as ``multiferroics", which display
simultaneously electric and magnetic ordering.
Most of the hexagonal manganites are ferroelectric 
below
a very high temperature (T$_{c}\approx$ 900 K),
and order magnetically at a lower
temperature (T$_{N}\approx$ 100 K).  A complex phase diagram
involving different magnetic 
order parameters has been
investigated using 
second harmonic generation \cite{Frolich,Fiebig1,Fiebig},
neutron scattering \cite{Lonkai} and heat capacity measurements \cite{Lorenz}.
Perhaps the most intriguing and technologically promising
development is the recent
observation of the 
magnetoelectric effect in the low
temperature region of the phase diagram of HoMnO$_3$ \cite{Lottermoser,Fiebig2}.
There are also indications of strong magneto-elastic coupling
in HoMnO$_3$ \cite{Cruz}.    

The manganese and rare earth spins in RMnO$_3$
are nearly geometrically frustrated
because of a slightly imperfect triangular lattice structure,
however, complications often associated with frustration,
such as spin liquid behaviour, are absent here.
Each magnetic phase is associated with a well-defined 
non-collinear spin structure within the hexagonal plane or
antiferromagnetic ordering along the $c$-axis.
The various magnetic phases have very similar antiferromagnetic (AF) 
interaction energies, due to the almost perfect
triangular lattice, which results in a close competition between them.

In each RMnO$_3$, the magnetic phase `$B_2$'
appears in zero magnetic field
at the temperature T$_N$, 
while the phase `$A_2$'
develops in magnetic fields of the order of few Tesla. 
These two phases are associated with ordering of Mn$^{3+}$ spins within 
the hexagonal
plane and are separated by a broad region of hysteresis. 
Additional phases appear at low
temperatures in HoMnO$_{3}$ \cite{Fiebig7}. 
A sharp Mn spin-reorientation transition occurs at  T$_{SR}\approx$ 33 K,
which results in the appearance of a third magnetic phase `$B_1$' \cite{Lorenz1}.
Below 33 K, a new intermediate phase, which exhibits the magneto-electric
effect, has been found in the region between 
the $B_1$ and $B_2$ phases \cite{Lottermoser,Lottermosernature}.
Moreover,  a fourth magnetic phase `$A_1$'
has been observed below 4 K due to
ordering of Ho$^{3+}$ spins \cite{Vajk}.

In Section II we describe the spin structures associated
with each magnetic phase, and the antiferromagnetic 
competition between them.
Section III we present a Landau model \cite{us} which describes the four magnetic
phases seen experimentally and numerical simulations
of phase diagrams derived from the model.
In Section IV we discuss our results.

\section{Crystal and Magnetic Structure\label{sec-structure}}

\begin{figure}[ht]
\centering
\subfigure[]
{\includegraphics[width=7.6cm]{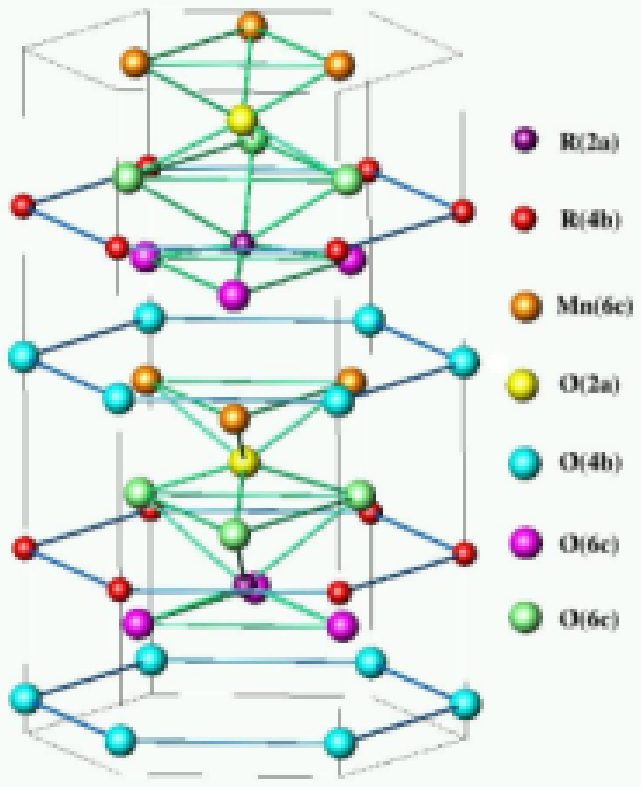}}
\subfigure[]{
\includegraphics[width=7.6cm]{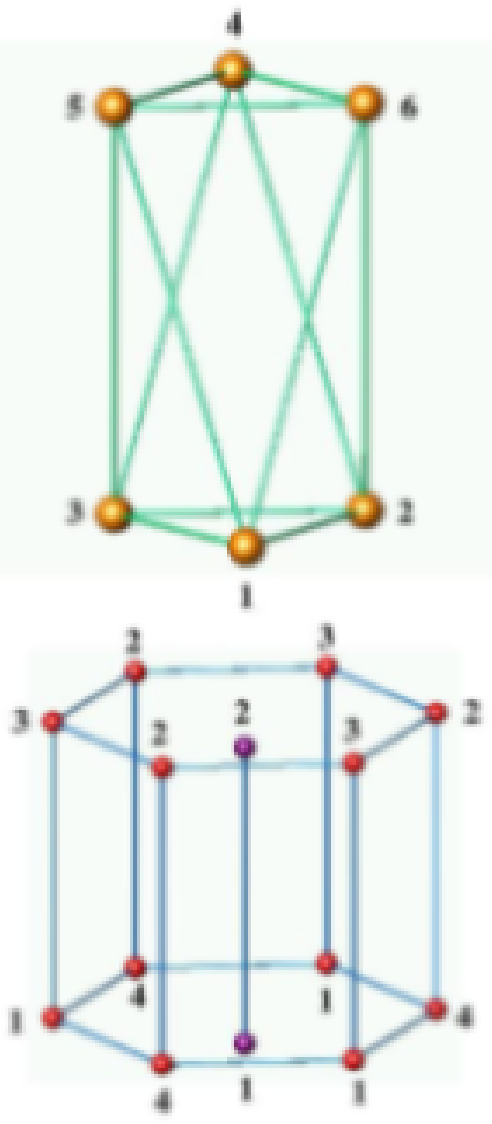}
}
\caption{a) Atomic positions in a single hexagonal 
primitive cell of  RMnO${_3}$. b)  Numbered Mn ions
at the (6c) positions (top)
and rare earth ions at the 
(4b) and (2a) positions (bottom). 
\label{fig-structure}}
\end{figure}

\begin{figure}[ht]
\centering
\subfigure[]{
\includegraphics[width=6.9cm]{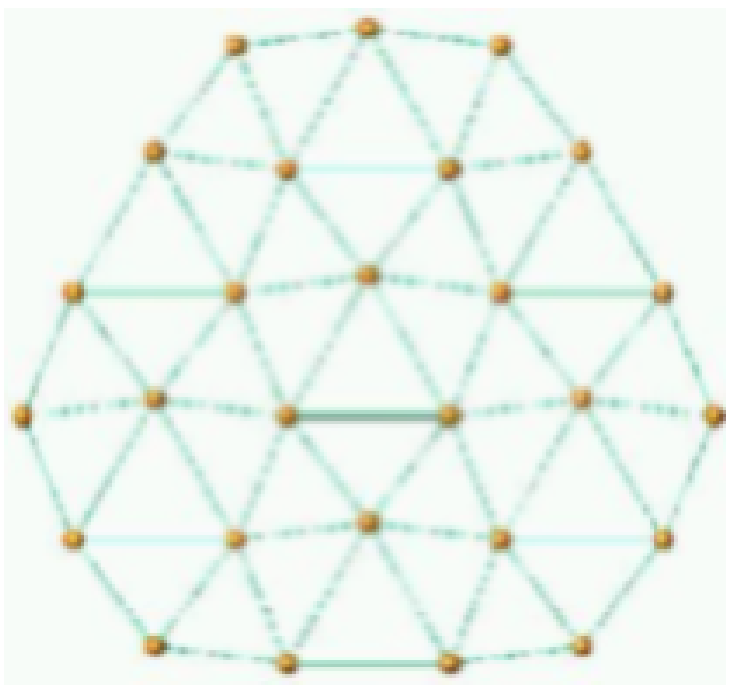}
}
\subfigure[]{
\includegraphics[width=8.1cm]{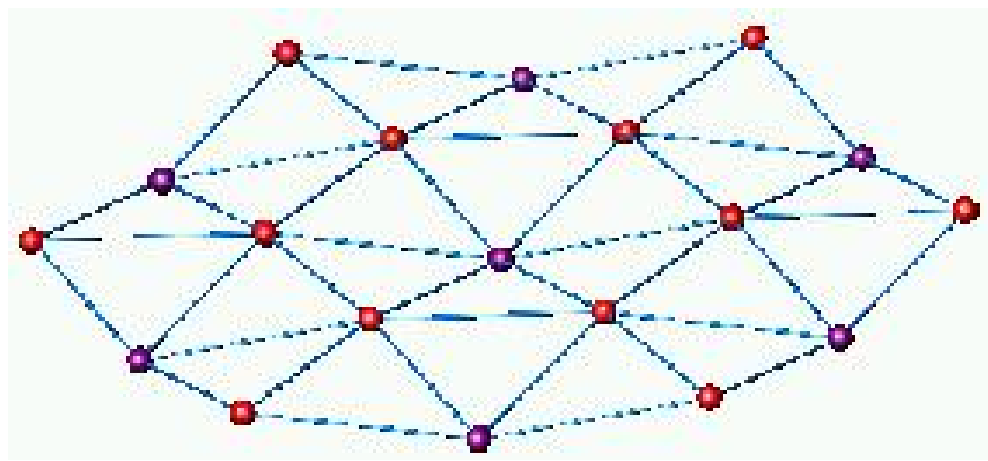}
}

\caption{Triangular lattices formed by a) Mn ions (viewed
parallel to the $c$-axis) and b)  rare earth ions (viewed at an angle
with respect to the $c$-axis).
\label{fig-triangles}}
\end{figure}

Below T$_{c}$, hexagonal RMnO$_3$ has the space group symmetry
P6$_3$cm (\#185, C$_{6v}^{3}$) \cite{Yakel}.
The crystal structure of RMnO$_3$ is shown in Figure \ref{fig-structure}a).
With six copies of the chemical
formula per unit cell, the Mn$^{3+}$ ions occupy the
(6c) positions,  
and form  triangular lattices on the $z$ and $z+1/2$ planes. 
The (6c) positions are $(x,0,z)$ and equivalent, where
$x\approx  a/3$ and $z = 0 $.  
$x=a/3$ yields a perfect triangular lattice.  
The rare earth ions occupy (2a) and (4b) positions, which are
$(0,0,z)$ (and equivalent, with
$z=0.22c$) and $(1/3,2/3,z)$ (and equivalent, with $z=0.27c$)
respectively.  A perfect triangular lattice is 
formed when the two $z$-parameters are equal.
The triangular lattices formed by the Mn and rare earth ions
are shown in Figure \ref{fig-triangles}.

The point group  C$_{6v}$ has
four one-dimensional irreducible representations (IR's), A$_{1}$, A$_2$, B$_{1}$ and
B$_2$, and two two-dimensional IR's E$_{1}$ and E$_2$. 
The same notation is used to label magnetic representations, 
where characters of `-1' in the
character table of C$_{6v}$ indicate a combination of the 
point group element with time reversal \cite{tinkham}.
For the 1D representations, the same names are given to the 
corresponding magnetic phases.
In the presence of magnetic order parameters,
the magnetic space groups are
P6$_3$cm (A$_1$), P6$_3$\b{c}\b{m} (A$_2$),
P\b{6}$_3$c\b{m} (B$_1$), and P\b{6}$_3$\b{c}m (B$_2$).

The spin configurations of rare-earth and Mn ions
may be classified according to the 
magnetic representations by which they transform.
All configurations which transform
according to 1D representations are
listed in Table~\ref{spins}, where the spin subscripts
refer to the atom numbers shown in Figure \ref{fig-structure}b).
The remaining degrees of freedom are accommodated by the
2D representations of C$_{6v}$;  but so far, there is no evidence that phases
corresponding to 2D OP's
appear in the phase diagram, so the 2D spin 
configurations are not included here.

\begin{table}[ht]
\begin{tabular}{ccl}
\hline
R (2a) & 
A$_2 \; \;$ & $S_{1z}+S_{2z}$\\
& B$_1$ & $S_{1z}-S_{2z}$\\
\hline
R (4b) & 
A$_1$ & $S_{1z}+S_{2z}-S_{3z}-S_{4z}$\\
& A$_2$ & $S_{1z}+S_{2z}+S_{3z}+S_{4z}$\\
& B$_1$ & $S_{1z}-S_{2z}-S_{3z}+S_{4z}$\\
& B$_2$ & $S_{1z}-S_{2z}+S_{3z}-S_{4z}$\\
\hline
Mn (6c)
& A$_1$ & $-(S_{1x}-S_{4x}) +\frac{1}{2}(S_{2x}+S_{3x}-S_{5x}-S_{6x})$\\
& & $
+\frac{\sqrt{3}}{2}(-S_{2y}+S_{3y}+S_{5y}-S_{6y}) $ \\
& A$_2$ & $-(S_{1y}-S_{4y}) +\frac{\sqrt{3}}{2}(S_{2x}-S_{3x}-S_{5x}+S_{6x})$\\
& & $
+\frac{1}{2}(S_{2y}+S_{3y}-S_{5y}-S_{6y}) $ \\
& B$_1$ & $-(S_{1y}+S_{4y}) +\frac{\sqrt{3}}{2}(S_{2x}-S_{3x}+S_{5x}-S_{6x})$\\
& & $
+\frac{1}{2}(S_{2y}+S_{3y}+S_{5y}+S_{6y}) $ \\
& B$_2$ & $-(S_{1x}+S_{4x}) +\frac{1}{2}(S_{2x}+S_{3x}+S_{5x}+S_{6x})$\\
& & $
+\frac{\sqrt{3}}{2}(-S_{2y}+S_{3y}-S_{5y}+S_{6y}) $ \\
& A$_2$ & $S_{1z}+S_{2z}+S_{3z}+S_{4z}+S_{5z}+S_{6z}  $\\
& B$_1$ & $S_{1z}+S_{2z}+S_{3z}-S_{4z}-S_{5z}-S_{6z}  $\\
\hline
\end{tabular}
\caption{1D spin configurations classified by irreducible representation.
The rare earth spin configurations are shown in Figure \ref{fig-hospins}
and the first four Mn (6c) configurations are shown in Figure \ref{fig-spins}.
\label{spins}}
\end{table}

We now discuss general features of the H-T phase
diagrams of RMnO$_3$ by considering
the relative strength of the nearest-neighbour antiferromagnetic interaction
for each spin configuration in Table \ref{spins}.
The antiferromagnetic interaction is defined as 
\begin{equation}
I = \frac{1}{2} J\sum_{<ij>} \hat S_i\cdot \hat S_j
\label{AF}
\end{equation}
where the sum is over nearest neighbours and $\hat S_i$ is a spin operator.
The 
parameter $J$ depends on the distance between
nearest neighbours.

\begin{figure}[ht]
\centering
\includegraphics[width=9cm]{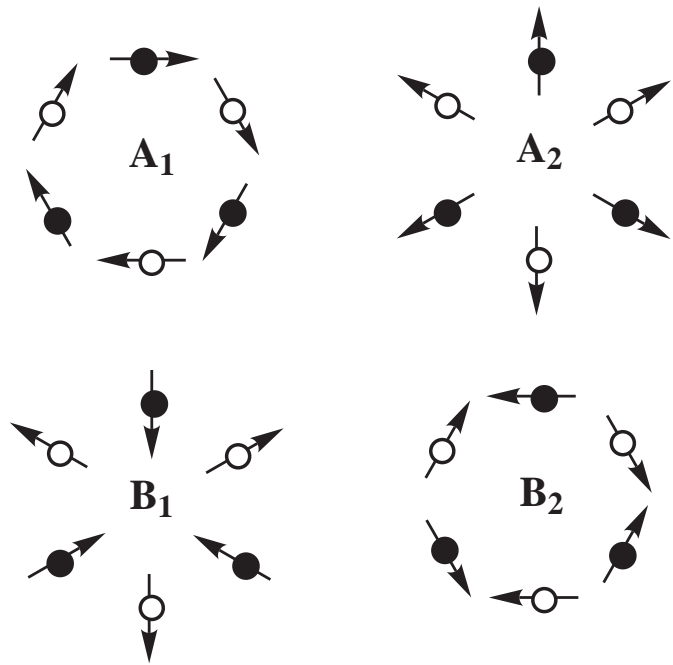}
\caption{Manganese spin configurations in the hexagonal plane.  These
correspond to the first four (6c) configurations listed in Table \ref{spins}\
(after  \cite{Fiebig}).
\label{fig-spins}}
\end{figure}

The first four configurations for Mn ions  listed in Table \ref{spins} and
illustrated in Figure \ref{fig-spins},
confine the spins to the hexagonal
plane. The $B_2$ phase that appears below 
$T_c$, the $A_2$ phase found in magnetic fields,
and the $B_1$ phase found at lower temperature in HoMnO$_3$
are due to these configurations.
In order to compare the AF interaction strengths for the 
different configurations, 
first, sets of nearest neighbours should be separated into three
cases, according to symmetry.  
The first case is the set of co-planar nearest neighbours 
($\{1,2,3\}$ and $\{4,5,6\}$, numbered as in Figure \ref{fig-structure}b)).  
In this case, the AF interaction is the same
for all four configurations, $I=-Js^2$.  The second and
third cases involve non-coplanar pairs.  The second case pairs
ions on opposite sides of the hexagonal primitive cell
($\{1,4\}$, $\{2,5\}$, $\{3,6\}$) and the
third case is the remaining  pairs.  The second case
favours $A_1$ and $A_2$ equally, while the third case favours $B_1$ and
$B_2$ equally, with $I=\pm Js^2$ in both cases.  
The distance between partners in each pair for the second and third cases is 
exactly the same if the Mn position parameter $x$ is exactly $1/3$.
Deviations away from $x=1/3$ will favour either the
$A$ phases or the $B$ phases.
However, the observed behaviour is the opposite of what could be
expected.  At higher
temperatures (75 K) $x=0.338(1)$ \cite{Lonkai}, which brings
the second case pairs closer together and favours the $A$ phases, but the 
$B_2$ phase is observed.  At the lowest temperature (1.5 K) $x=0.330(1)$ the
$B$ phases are favoured but 
the $A_1$ phase is observed.
The subtle competition between all four phases 
which results in the dominance of the $B_2$ phase at high temperatures is
most likely to 
be resolved by the inclusion of 
other interactions, such as next-nearest-neighbour, or
interactions
with the rare-earth ions.

\begin{figure}[ht]
\centering
\subfigure[]{
\includegraphics[width=4.8cm]{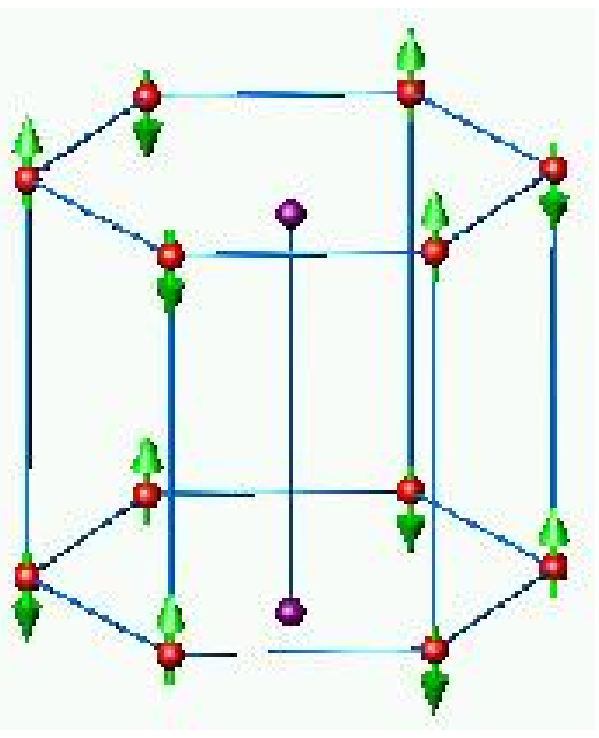}
}
\subfigure[]{
\includegraphics[width=4.8cm]{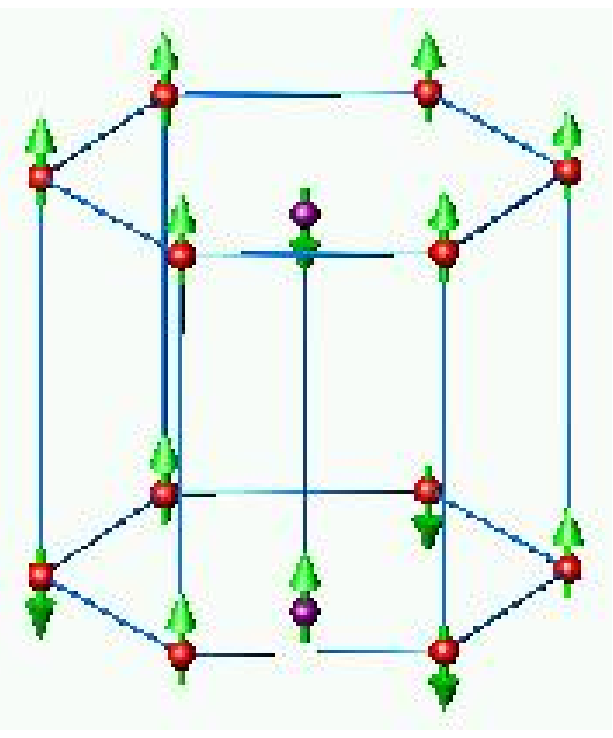}
}
\subfigure[]{
\includegraphics[width=4.8cm]{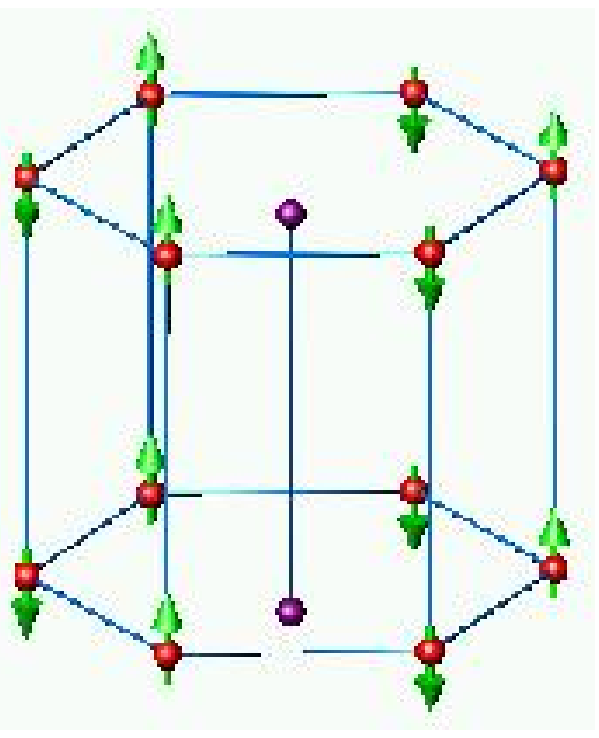}
}
\caption{Rare earth spin configurations  for  a) $A_1$,  b) $B_1$ and c) $B_2$
configurations for
(2a) and (4b) positions listed in Table~\ref{spins}.
\label{fig-hospins}}
\end{figure}

In HoMnO$_3$, additional phases $B_1$ followed by $A_1$  
appear as the temperature is further lowered.
The $B_1$ phase is associated both with in-plane Mn moments and
Ho ordering along the $c$-axis.  
The (2a) and (4b) holmium ions are almost co-planar.  
The $B_1$ phase is AF along the $c$-axis for (2a) and (4b) ions, but inside
the hexagonal plane, the (4b) ions are aligned ferromagnetically with each other, 
and antiferromagnetically with the (2a) ions.  This is shown in Figure \ref{fig-hospins}b).

At the lowest temperatures, the $A_1$ configuration emerges,
which has AF ordering of the (4b) spins in 
in the planes and along the 
$c$-axis, as shown in Figure \ref{fig-hospins}a).  Then, there can be no AF arrangement with respect to
the (2a) positions (because there is no $A_1$ configuration
for them).  
The phase transition from $B_1$ to $A_1$ is first-order,
as is the case between all transitions involving different
order parameters,  therefore hysteresis is anticipated, both
on phenomenological grounds, and also because of the 
persistence of the $B_1$ configuration on the (2a) positions.

The total in-plane AF interaction energy
for all Ho ions 
is approximately the same for all three configurations shown in Figure
\ref{fig-hospins}.
In all cases
$I=-2Js^2$, but $J$ differs according to the distance between ions.
If the (2a) and (4b) position parameters $z$ are equal then the three phases
are degenerate, otherwise $A_1$ and $B_2$ have lower energy.
The co-linear AF interactions favour the $B_1$ configuration, with $I=-3Js^2$
(versus $I=-2 Js^2$ for $A_1$ and $I=2 Js^2$ for $B_2$).


The $A_2$ phase will always dominate at
high enough magnetic fields, since it transforms in the same way as
the applied field, and couples linearly in the phenomenological sense.
Microscopically, Zeeman
coupling to Mn or Ho spins
induces ferromagnetic
order associated with the $A_2$ phase.

\section{Landau Model and Phase Diagram \label{sec-landau}}

The order parameters of the phases
A$_{1}$, A$_{2}$, B$_1$ and B$_{2}$ are denoted by $\eta_{1}$,
$\eta_2$, $\eta_3$ and $\eta_4$, respectively. 
The minimal Landau model which describes the A$_2$ and B$_2$ phases, observed
in all RMnO$_3$, is 
\begin{equation}
F = \alpha_2 \eta_2^2 + \beta_2 \eta_2^4 + \alpha_4 \eta_4^2 
+\beta_4 \eta_4^4 + \gamma_{24}\eta_2^2\eta_4^2
- H_z(\rho_1 \eta_2  +\rho_2 \eta_2^3 + \rho_3 \eta_2 \eta_4^2)
\label{24}
\end{equation}
where where $\alpha_i$, $\beta_i$ and $\gamma_{ij}$ and $\rho_i$
are phenomenological coupling constants and $H_z$ is the magnetic
field parallel to the $c$-axis.
The coefficients $\alpha_i $ are temperature dependent, $\alpha_i=a_i(T-T_i)$,
where $T_i$ is the temperature limit of stability for each phase
(which for convenience, we call the ``transition temperature"), and
$\beta_i > 0$
is required for stability.
$\alpha_4$ changes sign at $T_N \approx 100$K in all RMnO$_3$.

In zero applied field, the model allows for four different phases:
$(0,0)$ (the parent phase),
which may be connected to either $(\eta_2,0)$ (A$_2$-phase), $(0,\eta_4)$ (B$_2$-phase)
or $(\eta_2,\eta_4)$ (mixed phase) by second order phase transitions.
These are found by solving the set of coupled equations
$\partial F/\partial \eta_i = 0$, subject to the
minimisation
conditions $(\partial^2 F/\partial \eta_2^2)>0$ and $(\partial^2 F/\partial \eta_2^2)(\partial^2F/\partial\eta_4^2)
-(\partial^2F/\partial\eta_2\partial\eta_4)^2>0$.
The model also allows for the coexistence of two or more
different phases by hysteresis.

The mixed phase $(\eta_2,\eta_4)$ can be a
minimum of $F$ only when $4\beta_2\beta_4>\gamma_{24}^2$.  Its
existence is {\em not}
the result of hysteresis.
Anomalies in the $c$-axis magnetisation at the B$_2$ phase
boundary \cite{Lorenz}  are evidence that $B_2$ and $A_2$ are coupled
({\em i.e.} $\gamma_{24}\neq 0$).
In general,  $\eta_2$   grows linearly with applied field
but it may still be subject to a transition
in the sense that a change in sign of $\alpha_2$ will increase the number
of minima of the Landau functional.

Additional order parameters $\eta_1$ and $\eta_3$ are required to
describe the additional phases observed
in HoMnO$_3$.   Additional terms in the free energy include 
those 
obtained  by replacing in  (\ref{24}) $\eta_4^2$ by $\eta_1^2$ and $\eta_3^2$, 
as well as terms of the form $\eta_1\eta_2\eta_3\eta_4$
and $H_z\eta_1
\eta_3\eta_4$.

The Landau model describing A$_1$, A$_2$, B$_1$ and B$_2$ phases is 
\begin{eqnarray}
F(\eta_1, \eta_2, \eta_3, \eta_4)& = & F(\eta_2, \eta_4) + \alpha_1 \eta_1^2 + \beta_1 \eta_1^4 + \alpha_3 \eta_3^2 
+\beta_3 \eta_3^4 + \gamma_{12}\eta_1^2\eta_2^2 \nonumber \\
& & + \gamma_{13}\eta_1^2\eta_3^2 
 + \gamma_{14}\eta_1^2\eta_4^2 + \gamma_{23}\eta_2^2\eta_3^2 
  + \gamma_{34}\eta_3^2\eta_4^2 +\gamma \eta_1\eta_2\eta_3\eta_4  \nonumber \\
& & 
- H_z (\rho_4 \eta_2 \eta_3^2  + \rho_5 \eta_2 \eta_1^2
+\rho_6 \eta_1\eta_3\eta_4)
\label{1234}
\end{eqnarray}
This model may be solved exactly in zero applied field, and it
is found that the allowed phases are $(0,0,0,0)$ (the parent phase),
$(\eta_1,0,0,0)$ {\em etc.} (A$_i$ or B$_i$) and $(\eta_1,\eta_2,0,0)$ {\em etc.}
(mixed phases involving two order parameters).
In addition, two or more phases may co-exist due to hysteresis. 

The models (\ref{24}) and (\ref{1234}) were analysed by varying the 
temperature and field and searching for the minima of $F$ numerically.
Typically, several minima were present, the deepest corresponding
to the true ground state and the rest to metastable states
observed as hysteresis.  Each set of parameters $\alpha$, $\beta$,
$\gamma$ and $\rho$ yields a different phase diagram; these 
parameters were varied to find the best match to
phase diagrams obtained in experiments \cite{Fiebig,Lorenz,Vajk,Yen}.

\begin{figure}[ht]
\centering
\subfigure[]
{\includegraphics[width=7.7cm]{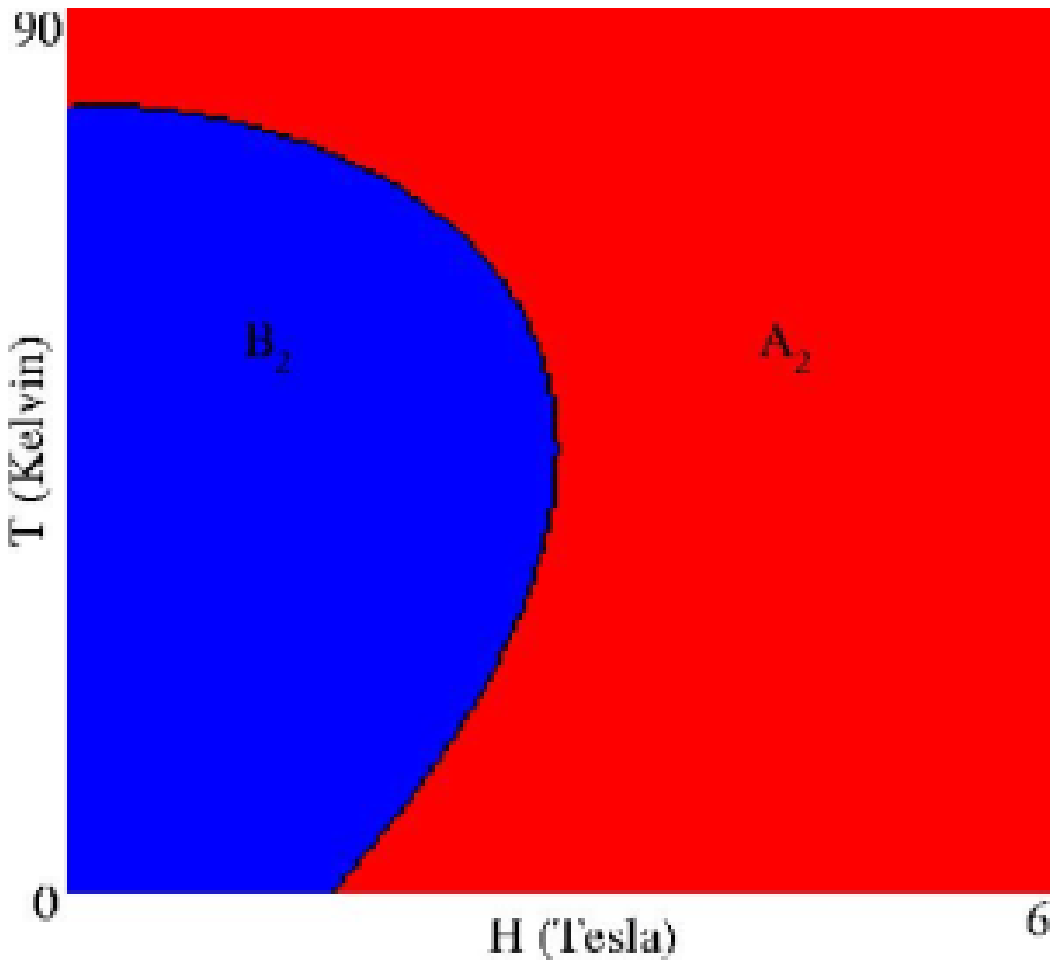}}
\subfigure[]
{\includegraphics[width=7.7cm]{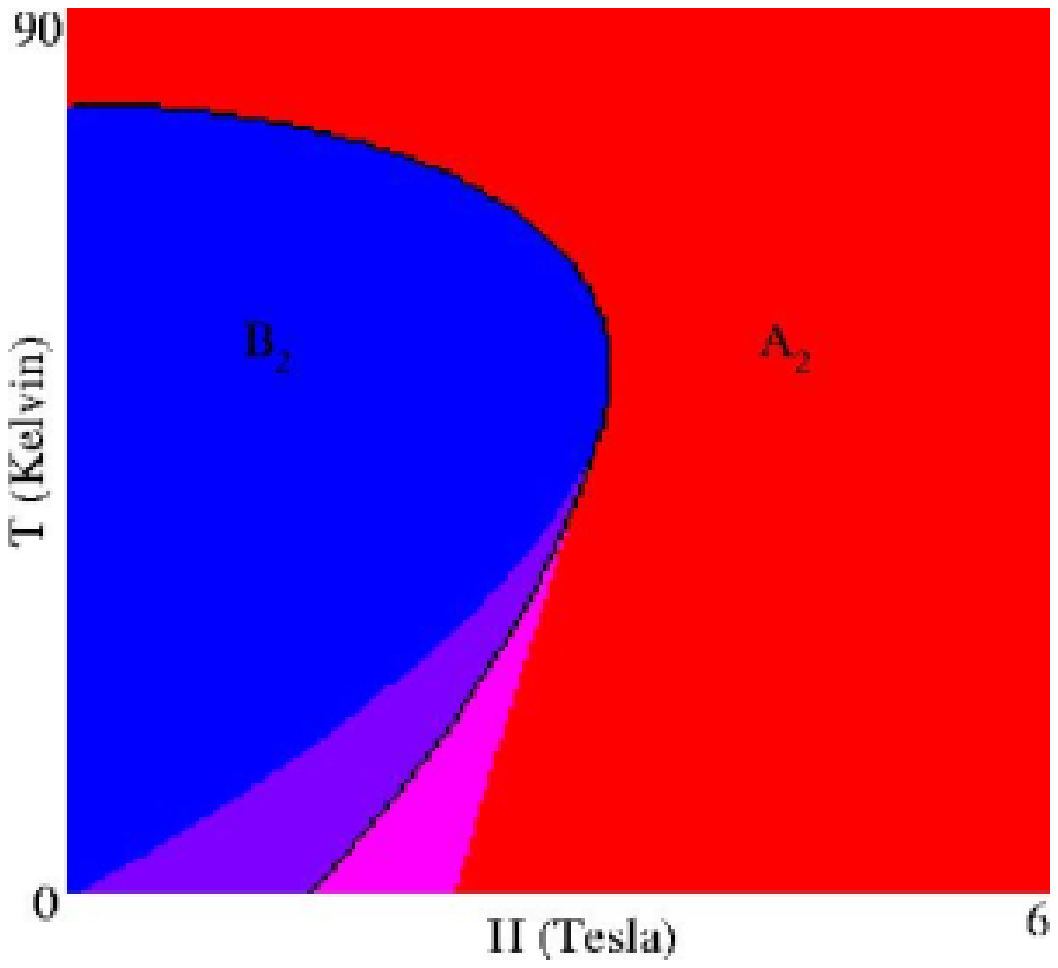}}
\\
\subfigure[]
{\includegraphics[width=7.7cm]{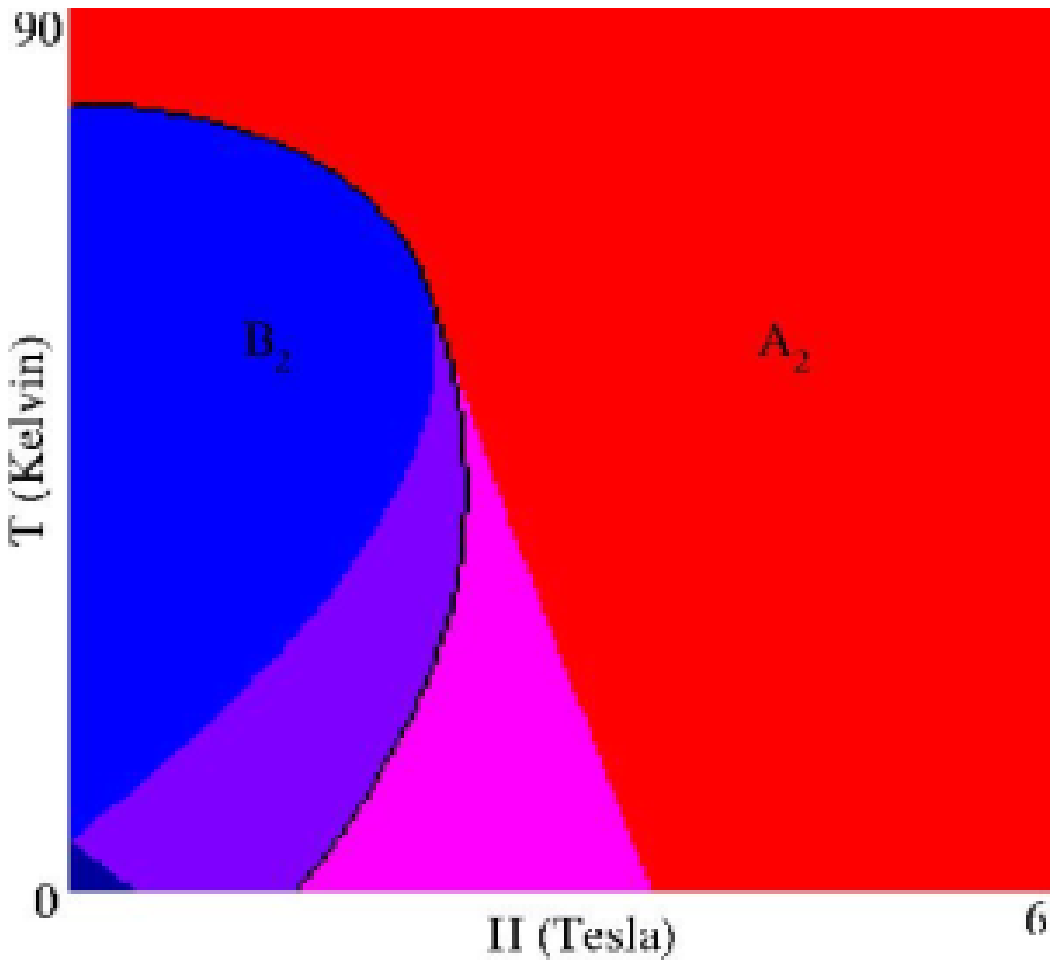}}
\subfigure[]
{\includegraphics[width=7.7cm]{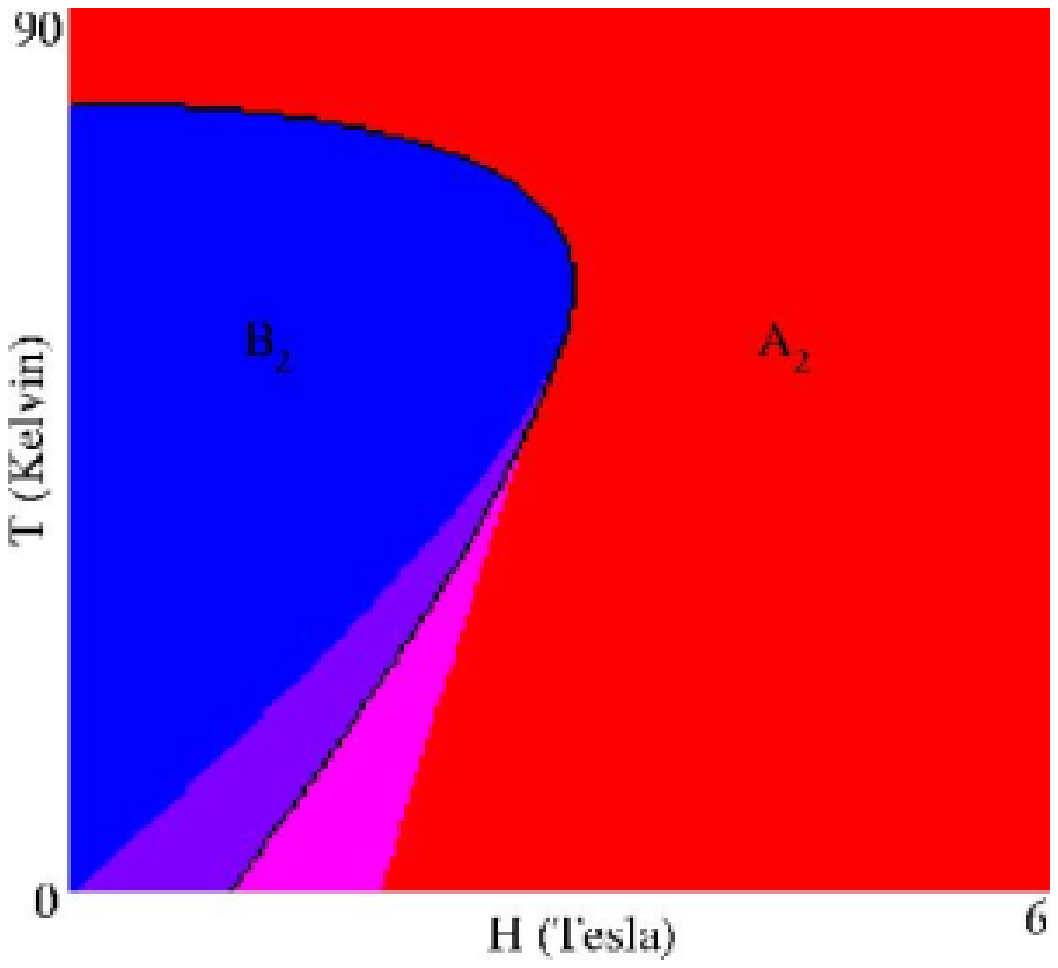}}

\caption{Numerical simulations of the phase diagram of  RMnO${_3}$.   The
black line is the boundary between
the $A_2$ and $B_2$ phases.  Hysteresis occurs in the lighter-coloured
regions.   In all four diagrams 
$T_4 = 80$~K, $T_2=10$ K,  $\beta_2=10$, $\beta_4=100$ and $\rho_1=12$, 
while $\gamma_{24}=60$, 150, 300 and 150 in 
(a), (b), (c) and (d) respectively. 
The parameters $a_i$ ($\alpha_i = a_i (T-T_i)$)
were scaled to unity.
 Nonlinear field-dependent 
coefficients $\rho_2 = \rho_3=5$ are introduced in diagram (d).
\label{fig-rmno3}} 
\end{figure}

Figure \ref{fig-rmno3} shows numerical simulations
of the phase diagrams for RMnO$_3$, modeled by (\ref{24}).
In all four diagrams, the onset of the $B_2$ phase for $H=0$ is
determined by setting $T_4=80$ K.  
Below this temperature two minima corresponding to 
non-zero $\eta_4$ occur in (\ref{24}). 
The $A_2$ phase, which dominates the right side (high field region) of 
all diagrams, corresponds to a Landau functional which has
only {\em one} minimum that is shifted away from $\eta_2 = 0$
because of the term $H_z\eta_2$.  Thus the free energy resembles
the parent phase but the symmetry is
$A_2$.
In Figures \ref{fig-rmno3}b), c) and d)
hysteresis occurs, indicated by lighter coloured areas near the black phase
boundary line. 
Here the free energy has three minima,
one corresponding to the field-shifted parent phase, and two to the $B_2$ phase.
In Figures \ref{fig-rmno3}a), b) and d) we have 
$\gamma_{24}^2 < 4\beta_2\beta_4$, so
a pair of minima for $\eta_2\neq 0$ never occurs.  
However, in  Figure \ref{fig-rmno3}c), $\gamma_{24}^2 > 4\beta_2\beta_4$, 
so two shallow minima for $\eta_2$ co-exist with two for $\eta_4$ in the 
bottom left corner of the phase diagram.
Figure \ref{fig-rmno3}b) shows the 
correct arrangement of phases on the phase diagram compared to 
experiments, but fails to 
simulate correctly the curvature of the phase boundary.
Fig ure~\ref{fig-rmno3}d), which
includes the non-linear (in OP) field-dependent terms in (\ref{24}),
is a significant improvement.

\begin{figure}[ht]
\subfigure[]{
\includegraphics[width=7.5cm]{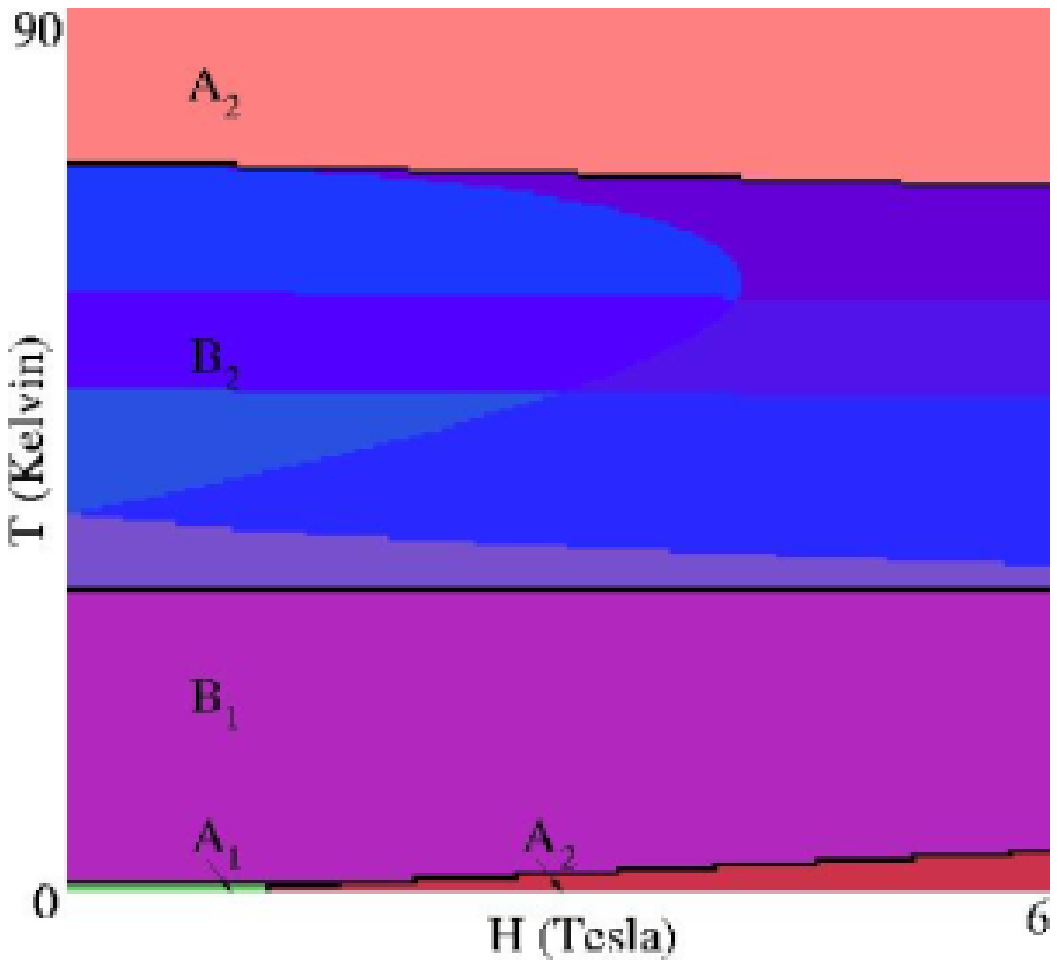}
}
\subfigure[]{
\includegraphics[width=7.5cm]{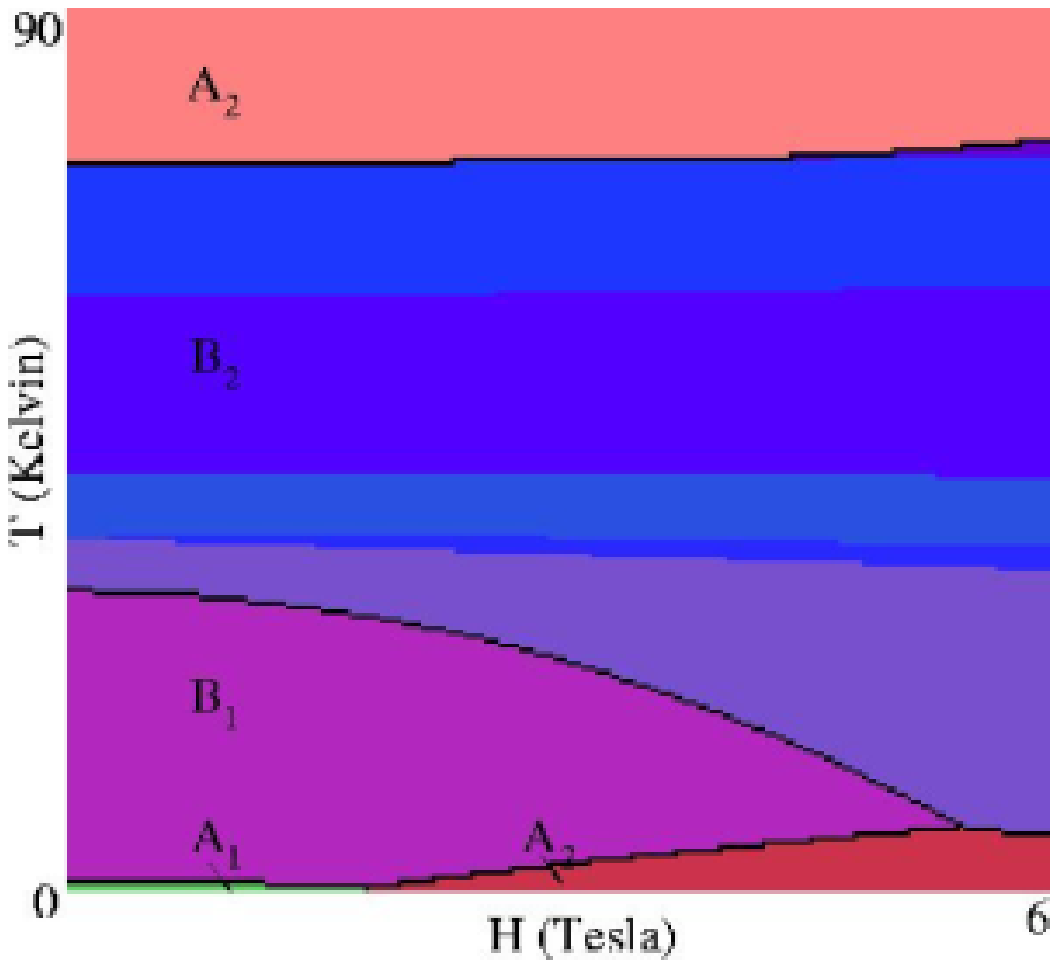}
}
\caption{Numerical simulations of the phase diagram of HoMnO${_3}$.
Black lines separate the phases, while different shades
within a phase represent the presence of other meta-stable states.
In both diagrams $\beta_1 = 0.14$, $\beta_2=0.08$, $\beta_3=0.2$, $\beta_4=0.4$,
while $\gamma_{12}=1$, $\gamma_{13}=1$, $\gamma_{14}=1$, $\gamma_{23}=6$,
$\gamma_{24}=2.9$, $\gamma_{34}=10$, $\rho_1=1$ in (a) and
$\gamma_{12}=90$, $\gamma_{13}=\gamma_{14}=\gamma_{23}=  
\gamma_{24}=\gamma_{34}=100$, $\rho_1=4$ in (b).
Diagram (b) also
includes non-linear field-dependent terms $\rho_2 = 1$, $\rho_3=2$,
$\rho_4 = 0.05$.
\label{fig-homno3}
} 
\end{figure}

Figure \ref{fig-homno3} shows numerical simulations of the phase
diagram of HoMnO$_3$, modeled by  (\ref{1234}).
The upper part of the diagrams, showing the $B_2$ and $A_2$ phases, 
is similar to RMnO$_3$, in that the $B_2$ phase appears below
the transition temperature $T_4 = 75$ K, and the $A_2$ phase is induced by the
magnetic field.  Non-zero transition temperatures $T_i$ were assigned to all
other phases ($T_1=52$ K, $T_2=39$ K, $T_3 =62$ K)
below which the phases are metastable and
hysteresis occurs.  True phase
transitions occur below the transition temperatures, along the
lines where the free energies, evaluated for different phases, are equal.
Different shades within a single phase represent the presence of other,
metastable phases.  
Figure \ref{fig-homno3}a) bears a poor resemblance to experiments.  Figure
\ref{fig-homno3}b), which includes non-linear field-dependent
terms, is significantly better,  but still fails to reproduce qualitatively
all of the phase boundary around the $B_1$ phase  as seen
in experiments.

\section{Discussion}

In Section \ref{sec-structure} we argued that even in the absence of
true geometrical frustration, there is not enough information
to predict the magnetic ground state without detailed knowledge
of the AF interaction strength $J$. 
In Landau theory, the microscopic model (\ref{AF}) is replaced by a
phenomenological one (\ref{24},\ref{1234}), and the parameter $J$
is incorporated into the temperature-dependent $\alpha_i$.
Landau theory includes all interactions allowed by symmetry, and as such is
more general than the AF interaction, which is isotropic.  However, the
proliferation of phenomenological constants also inhibits the 
predictive powers of the Landau
model.   Nevertheless,
Landau modeling is useful because it can reveal the minimal elements
in a theory 
that are needed to describe the phase diagrams of  
RMnO$_3$.   
In our analysis, we found that
a model based on the usual second and fourth order terms, and a linear 
coupling of the order parameter to the magnetic field, does not describe well
the observed phase diagrams, especially the curvature of the phase boundaries.
The inclusion of non-linear (in OP) field-dependent terms 
is a significant improvement.

The magneto-electric effect is observed in the region between the $B_1$ and
$B_2$ phases.  Linear coupling of the magnetic and electric
fields of the form $\alpha_{ij}E_iB_j$ can only occur when both
inversion and time reversal symmetry are absent - these are necessary but not
sufficient conditions.   A mixture of $B_1$ and $B_2$ order parameters (due
to hysteresis)
does not lower the symmetry enough for magneto-electric coupling.
However, domain walls, which connect different domains of the same phase,
have been implicated in the observation of the 
magneto-electric effect \cite{Lottermoser}.
Thus OP gradient terms, which couple to the magnetic field,  may 
significantly alter the free energy landscape, and could possibly replace
the non-linear field-dependent terms which we introduced.

In conclusion,
we have studied phase diagrams for RMnO$_3$, using group theory and 
Landau theory, by 
including up to fourth order
phenomenological couplings between order parameters and non-linear
coupling to 
an applied magnetic field.  Antiferromagnetic 
competition between magnetic phases, due to a 
near perfect triangular lattice
structure,
gives rise to a complex phase diagram.
Our simulations reproduce the main features seen in experimental
results, including the general arrangement of phases on the diagram,
hysteresis effects and the curvature of phase boundaries.

\ack
We thank I. Sergienko for many helpful discussions.  
This work was supported by NSERC of Canada.

\end{document}